\documentclass[conference]{IEEEtran}
\IEEEoverridecommandlockouts
\usepackage{cite}
\usepackage{amsmath,amssymb,amsfonts}
\usepackage{algorithmic}
\usepackage{graphicx}
\usepackage{multirow}
\usepackage{booktabs}
\usepackage{makecell}
\usepackage{textcomp}
\usepackage[table]{xcolor} % Allows coloring cells
\usepackage{colortbl} % Additional color tools
\usepackage{amssymb}
\usepackage{booktabs}
\usepackage{arydshln}
\usepackage[margin=0.5in]{geometry}
\usepackage{multirow}
\usepackage{graphicx}
\usepackage{array}
\usepackage{makecell}
\usepackage{relsize}

\def\BibTeX{{\rm B\kern-.05em{\sc i\kern-.025em b}\kern-.08em
    T\kern-.1667em\lower.7ex\hbox{E}\kern-.125emX}}

\usepackage{arydshln}

\usepackage[status=final,nomargin,inline]{fixme}
\usepackage{placeins}
\usepackage{xcolor}
\fxusetheme{colorsig}
\FXRegisterAuthor{dg}{adg}{\textcolor{green}{by DG}}
\FXRegisterAuthor{td}{atd}{\textcolor{green}{ToDo}}
\FXRegisterAuthor{hw}{ahw}{\textcolor{red}{for HW Team}}
\FXRegisterAuthor{ma}{mag}{\textcolor{blue}{MA}}
\FXRegisterAuthor{bh}{bhg}{\textcolor{purple}{by BH}}
\FXRegisterAuthor{ep}{epg}{\textcolor{cyan}{EP}}
\FXRegisterAuthor{gk}{gkg}{\textcolor{magenta}{by GK}}

\usepackage[hyphens]{url} % For line breaks in URLs
\usepackage{xurl}         % Smarter URL breaking

\usepackage{xspace}
\usepackage{textcase}

\newcommand{\smalltextsc}[1]{\textsc{\small #1}}
\newcommand{\passmetric}[1]{\smalltextsc{Pass@}#1\xspace}

\newcommand{\turtle}{\smalltextsc{TuRTLe}\xspace}

\usepackage[colorlinks=true, linkcolor=red, citecolor=blue, urlcolor=magenta]{hyperref}

\usepackage{textcomp}
\usepackage{amssymb}

\newcommand{\eg}{\emph{e.g.}, }       % for example
\newcommand{\bigcode}{\texttt{BigCode Evaluation Harness}}

\newcommand{\leaderboardurl}{\texttt{https://huggingface.co/spaces/HPAI-BSC/TuRTLe-Leaderboard}}
\newcommand{\githuburl}{\texttt{https://github.com/HPAI-BSC/TuRTLe}}
\newcommand{\redesurl}{\texttt{https://www.red.es/es}}

\makeatletter
\newcommand{\linebreakand}{
  \end{@IEEEauthorhalign}
  \hfill\mbox{}\par
  \mbox{}\hfill\begin{@IEEEauthorhalign}
}
\makeatother

\usepackage{soul}

\begin{document}

%TuRTLe: A Unified Evaluation for RTL Generation
\title{TuRTLe: A Unified Evaluation of LLMs for RTL Generation}

\author{
\IEEEauthorblockN{Dario Garcia-Gasulla}
\IEEEauthorblockA{\textit{Barcelona Supercomputing Center} \\
dario.garcia@bsc.es}
\and
\IEEEauthorblockN{Gokcen Kestor}
\IEEEauthorblockA{\textit{Barcelona Supercomputing Center} \\
gokcen.kestor@bsc.es}
\and
\IEEEauthorblockN{Emanuele Parisi}
\IEEEauthorblockA{\textit{Barcelona Supercomputing Center} \\
emanuele.parisi@bsc.es}
\linebreakand
\IEEEauthorblockN{Miquel Albertí-Binimelis}
\IEEEauthorblockA{\textit{Barcelona Supercomputing Center} \\
miquel.alberti@bsc.es}
\and
\IEEEauthorblockN{Cristian Gutierrez}
\IEEEauthorblockA{\textit{Barcelona Supercomputing Center} \\
cristian.gutierrez@bsc.es}
\and
\IEEEauthorblockN{Razine Moundir Ghorab}
\IEEEauthorblockA{\textit{Barcelona Supercomputing Center} \\
moundir.ghorab@bsc.es}
\linebreakand
\IEEEauthorblockN{Orlando Montenegro}
\IEEEauthorblockA{\textit{Barcelona Supercomputing Center} \\
orlando.montenegro@bsc.es}
\and
\IEEEauthorblockN{Bernat Homs}
\IEEEauthorblockA{\textit{Barcelona Supercomputing Center} \\
bhomsgis@bsc.es}
\and
\IEEEauthorblockN{Miquel Moreto}
\IEEEauthorblockA{\textit{Barcelona Supercomputing Center} \\
\textit{Universitat Politecnica de Catalunya}\\
miquel.moreto@bsc.es}
}

\maketitle

\begin{abstract}
Rapid advancements in LLMs have driven the adoption of generative AI in domains like Electronic Design Automation (EDA). Within the field of software development, EDA presents unique challenges derived from specific requirements of generated RTL code; RTL code must not only be syntactically correct and functionally accurate, but also synthesizable by hardware generators, while matching performance, power and area (PPA) constraints. These additional requirements introduce complexities that existing code-generation benchmarks often fail to capture, limiting their effectiveness in evaluating LLMs for RTL generation.
To address this gap, we propose \turtle{}, a unified evaluation framework designed to systematically assess LLMs across key RTL generation tasks. \turtle{} integrates multiple existing benchmarks and automates the evaluation process, enabling a comprehensive assessment of LLM performance in syntax correctness, functional correctness, synthesis, PPA optimization, and exact line completion. Using this framework, a diverse set of forty open LLMs are assesed, tracking their strengths and weaknesses in EDA-specific tasks.
Our results identify the best match for specific tasks (\textit{e.g.,} base models are better in module completion tasks, instruct-tuned models are better in specification-to-RTL tasks), while finding that recent models with autoregressive reasoning chain perform the best overall. Some benchmarks, particularly within syntax correctness, show signs of saturation, while others remain as open problems for LLMs.

%old version
%LLMs are ready to contribute to RTL code generation. To enable the development of robust models for that purpose, a reliable evaluation methodology needs to be established. Recent works have proposed several benchmarks, each covering different aspects of RTL code generation (syntax, functionality, synthesis, etc.), but these, individually, remain limited in size and scope. In this work, we release a unified evaluation framework, \turtle{}, which integrates a variety of existing RTL benchmarks through the scalable and extendable Evaluation Harness. In its first release, the goals covered by \turtle{} include syntax checking, functionality correctness, synthesis, PPA optimization and exact line completion. This is done by integrating and extending four different benchmarks, which combined produce a comparatively large sample size. After the definition of \turtle{}, a large set of open models are evaluated under it in a comprehensive assessment of RTL generators. Insights found regarding evaluation practices and model outcomes are discussed, answering questions such as: Which model is best for each task? Which benchmarks is harder to solve? How does performance across benchmarks correlate?
\end{abstract}

%\begin{IEEEkeywords}
%RTL generation, LLMs for EDA, Benchmarking
%\end{IEEEkeywords}

\section{Introduction}
\label{sec:intro}

Advancements in large language models (LLMs) have unlocked new possibilities across a wide range of domains~\cite{BrownNIPS20,Bubeck23}. Domain-specific LLMs have gained significant attention due to their strong performance in specialized tasks, including financial engineering~\cite{ShijieWu2023}, biomedical research~\cite{shin-etal-2020}, and scientific computing~\cite{taylor2022galactica, acharya2024}. In software related tasks, LLM can suggest code snippets, solve common coding challenges, and provide explanations of complex concepts~\cite{Mastropaolo2023,nijkamp2023codegen,roziere2024codellama,starchat2}.

In the field of Electronic Design Automation (EDA), researchers are increasingly exploring the use of LLMs to accelerate hardware design\cite{chen2024EDA, PanSurveyEDA25, NakkabMLCAD24}. In the traditional digital system design flow, engineers must implement precise functionality using hardware description languages (HDLs). LLM-based solutions aim to bridge this gap by translating functional specifications directly into HDL code, such as Verilog. This approach has the potential to revolutionize hardware design and verification by streamlining Verilog coding, optimizing circuit implementations, and automating time-consuming design tasks~\cite{ChipChat23}. Recent research has explored various techniques for Verilog code generation, including prompt engineering to enhance model responses~\cite{chang2023chipgpt}, training or fine-tuning on Verilog-specific datasets~\cite{thakur2024verigen, liu2024chipnemo, liu2024rtlcoder, ZehuaBetterV_24, dehaerne2023}, and
agent-based approaches to improve iterative refinement and debugging~\cite{TsaiRTLFixer24, huang2024VeriAssist, islam2024aivril, zhao2024mage, mi2024promptv}.
However, several research questions about the quality and fidelity of the generated Verilog code remain largely unanswered.
LLM-based solutions need to generate code that is syntactically and functionally correct, but in EDA the generated code must also be synthesizable to produce correct hardware components, and achieve reasonable performance in non-functional indicators such as 
area and power.
%It is evident that evaluation methodologies and benchmarks designed to assess LLM that produce traditional code do not fully cover the entire evaluation space and are not completely suitable to evaluate LLM-based EDA.

To mitigate this issue and assess the effectiveness of current LLM for EDA, several benchmarks have been introduced, including VerilogEval\cite{liu2023verilogeval,pinckney2025revisiting}, RTL-Repo\cite{allam2024rtl}, and RTLLM\cite{lu2024rtllm}. 
While these benchmarks evaluate different aspects of Verilog code generation, including single-line completion, single-module generation, and specification-to-RTL translation, none fully evaluates the entire hardware design flow, and some use different metrics. This makes it challenging to compare results across benchmarks and LLMs.
A standardized evaluation methodology and infrastructure would allow researchers to systematically assess LLM capabilities, identify strengths and limitations in existing benchmarks and models, and drive further advancements in LLM-driven hardware design automation.

This work introduces \turtle{}, a comprehensive evaluation framework for RTL generation. It integrates multiple benchmarks into a single fully-automated evaluation framework, providing a standardized and comprehensive assessment of code. Coverage is provided for different RTL design goals, including syntax and correctness of the generated code, synthesizability of the hardware circuits produced by hardware generators, together with their corresponding performance, area, and power metrics, using a novel metric based on comparison against human performance. These goals are tested under different conditions, such as module completion, and generation from specifications expressed in natural language. \turtle{} is used to benchmark forty models, providing a comprehensive, wide and updated review on the state of open models. This includes general purpose, coding, and RTL-specific LLMs, as well as base, instruct and reasoning models, showing the strengths and weaknesses of each alternative. 

Our analysis shows that temperature, context length and training strategy (\eg base, instruct, reasoning) all play a crucial role in LLM performance across EDA-related tasks. Base models demonstrate stronger performance in module completion, effectively filling in partial RTL definitions, while instruct-tuned models are better at generating RTL from natural language specifications. Models employing reasoning prompting chains, exhibit remarkable performance across the board at the cost of increased inference time and token generation. Overall, our findings indicate that while current LLMs are highly reliable in generating syntactically correct RTL code, they still struggle with functional correctness and in achieving human-level non-functional indicators.

\section{Evaluation Methodology}
\label{sec:evalmeth}
%\subsection{An overview of the Evaluation Framework}

Evaluating LLM-generated RTL code requires a comprehensive and systematic approach that considers the many aspects of the hardware design process. \turtle{} integrates existing benchmarks including VeriGen~\cite{thakur2024verigen}, VerilogEval~\cite{liu2023verilogeval,pinckney2025revisiting}, RTL-Repo~\cite{allam2024rtl}, and RTLLM~\cite{lu2024rtllm}, within a single infrastructure, allowing for direct comparison of LLM performance across various RTL generation tasks. Focusing on versatility and automation in the construction and running of \turtle, we choose a widely adopted framework in the field of code generation (\bigcode~\cite{bigcode}) as a foundation. 
This tool can easily load a variety of generative models, and already implements multiple LLM tasks and metrics. %Focusing on scalability, we fork from the \vllmharness{} project\footnote{https://github.com/iNeil77/vllm-code-harness}, a particularly efficient derivative, which facilitates running larger models through the vLLM~\cite{kwon2023efficient} library. 
%Through these choices, \turtle{} is compatible by design with most popular practices in the field. 

This section first reviews three tasks involved in RTL generation that \turtle{} focuses on: single-line completion, module completion, and specification-to-RTL conversion, all in \S\ref{subsec:tasks}. Next, it outlines the five RTL design goals assessed by \turtle{}: line-level contextual accuracy, syntax correctness, functional correctness, synthesizability, and post-synthesis quality (see \S\ref{subsec:goals}). To evaluate LLM performance on these goals, we employ several metrics, including exact matching, \passmetric{1}, and a novel PPA-Score (detailed in \S\ref{subsec:metrics}). Table~\ref{tab:metrics} shows the relationships between the various tasks, design goals, metrics, and tools used in the \turtle framework.

\subsection{Generation Tasks}
\label{subsec:tasks}

Our evaluation framework assesses LLMs across three fundamental RTL code generation tasks, each representing a distinct level of complexity and scale. The smallest of tasks, in the sense of generation length, is \textbf{Single-Line Completion (SLC)}, which focuses on the model’s ability to predict the next line of code given a partial context. This task closely resembles auto-completion scenarios, where engineers rely on coding assistants to streamline development. 

The second task in complexity is \textbf{Module Completion (MC)}, which requires LLMs to generate a complete module based on a given function description or module signature. This task evaluates an LLM's ability to process the behavioral description of a hardware module and produce correct implementations.
%This task evaluates an LLM’s ability to process hardware functionality, defining correct module structures and producing functionally accurate implementations.

Finally, in \textbf{Specification-to-RTL (S2R)} the LLM must generate a complete implementation from a natural language hardware specification. This task does not provide a predefined module interface, so models must infer additional information from the specification, such as the module name, the correct ports and parameter types. This mimics the process of translating human-readable specifications into working hardware descriptions. Given their fundamental differences in complexity and nature, results for these three approaches are reported separately.

\subsection{Design Goals}\label{subsec:goals}

\turtle defines five design goals to evaluate LLM-generated RTL code, ensuring a comprehensive assessment of correctness, feasibility, and efficiency within the hardware design workflow. These goals are aligned with the supported tasks as shown in Table~\ref{tab:metrics}.

%\textbf{Line-level Contextual Accuracy (LCA)} rates models based on their capacity at producing short code completions which are to be as close as possible to a given reference answer. This goal aligns entirely with the SLC task, and implicitly evaluates (through exact match with the golden solution) whether the predicted line is both syntactically correct and contextually coherent within the surrounding logic. 
\textbf{Line-level Contextual Accuracy (LCA)} measure the ability of the models to reproduce line completions in accordance to a given reference answer. This goal aligns with the SLC task, and partially evaluates whether the predicted line is both syntactically and functionally correct.

The four remaining goals can all be applied to both MC and S2R tasks. %\textbf{Syntax Correctness (STX)} ensures that the generated code adheres to HDL syntax rules and can be processed by standard compilers and simulators~\cite{tsai2024rtlfixer}. The validity of generated designs is verified by compiling the code alongside its testbench using a tool that checks for parsing errors, missing constructs, and incorrect syntax. 
\textbf{Syntax Correctness (STX)} ensures that the generated HDL code complies with the language’s grammar rules and can be processed by standard HDL analyzers, synthesizers, and simulators~\cite{tsai2024rtlfixer}. The correctness of generated designs is verified by tools that check for syntax errors, missing constructs, and structural inconsistencies.
\textbf{Functional Correctness (FNC)} ensures that the generated HDL code exhibits the expected behavior as specified in the design requirements (prompt)~\cite{liu2023your, pulavarthi2024assertionbench}. FNC is verified through behavioral simulation, where the generated design is executed alongside a predefined testbench in an HDL simulator. While this approach provides a practical verification method, it relies on the assumption that testbenches accurately capture functional discrepancies. 
%This assumption holds for small designs or well-defined arithmetic modules but can be a limitation for complex designs, highlighting the need for formal verification techniques in future work.
%\textbf{Functional Correctness (FNC)} evaluates whether the generated code matches the expected behavior described in the prompt for the problem description~\cite{liu2023your, pulavarthi2024assertionbench}. 
%EP: any problem with module interface and parameters is already spotted during compilation.
%This assessment ensures that the module interface and parameter configurations are correctly implemented and that the design functions as intended when simulated~\cite{liu2023your, pulavarthi2024assertionbench}. 

%FNC is verified through behavioral simulation using a compiler, where the generated code is tested against a predefined testbench. While this approach provides a practical verification method, it assumes that testbenches accurately capture any functional discrepancies. Even though it is a reasonable assumption for small designs or well-defined arithmetic modules, it might be a potential limitation for complex designs, making the adoption of formal methodologies desirable in future work.
\textbf{Synthesizability (SYN)} assesses whether the generated HDL code can be successfully synthesized into a gate-level netlist using a synthesis tool. In practical hardware design, these tools support only a subset of HDL constructs, making synthesizability a key criterion for real-world applicability. SYN is verified by processing the generated RTL code using a synthesis toolchain, ensuring that the design is implementable in hardware. This is critical for models that generate RTL intended for manufacturable chip designs.
%\textbf{Synthesizability (SYN)} assesses whether the generated code can be successfully synthesized into a gate-level netlist. In practical hardware design, only a subset of HDL constructs is synthesizable, making this goal essential for evaluating real-world applicability. SYN, which is verified by elaborating the generated RTL code through a synthesis framework, is critical for models that generate RTL for manufacturable chip designs. 

Finally, \textbf{Post-Synthesis Quality (PSQ) }evaluates the implementation efficiency of the synthesized design based on Power, Performance, and Area (PPA). PPA metrics are widely used in chip design optimization to compare different implementations and ensure that a design meets key hardware constraints~\cite{thorat2023advanced}. In this work, we use PPA to compare LLM-generated designs with human-crafted reference implementations. We set up a unified synthesis pipeline based on OpenLANE~\cite{shalan2020building} to extract PPA from pairs of implementations of the same specification, one generated by an LLM and one by a human. Then, we extract area and power directly from the PPA report and evaluate performance based on maximum delay, defined as the difference between the clock period and the worst slack reported by static timing analysis.% All three PPA metrics are represented as positive numbers, where $0$ represents the minimum possible value, leaving the maximum unbounded.
%\textbf{Post-Synthesis Quality (PSQ)} evaluates the hardware efficiency of the synthesized design in terms of Power, Performance, and Area (PPA). PPA metrics are widely used in chip design optimization to compare different implementations and ensure that a design meets non-functional requirements~\cite{thorat2023advanced}.
%In this work we use PPA to compare LLM-generated designs with human-crafted reference implementations. We extract area and power directly from the PPA report, and evaluate performance as the maximum delay of the design, computed as the difference between the clock period and the worst slack reported by the static timing analysis.% In this way, all three PPA metrics are represented as positive numbers, where $0$ represents the minimum possible value, leaving the maximum unbounded.

\begin{table}[t]
    \centering
    \caption{Design goals and metrics for benchmarking RTL generation. The use of open-source tools ensures accessibility and reproducibility.}
    \begin{tabular}{l|l|l|l}
        \textbf{Tasks} & \textbf{Goals} & \textbf{Metrics} & \textbf{Tools} \\
        \midrule
        SLC & LCA & \multirow{4}{*}{\passmetric{$1$}} & Exact Match \\
        \cline{1-2} \cline{4-4}
        \multirow{4}{*}{MC} & STX &  & \multirow{2}{*}{Icarus Verilog}\\
        \cline{2-2}
        \multirow{4}{*}{S2R}
                          & FNC  &  & \\
        \cline{2-2} \cline{4-4}
                          & SYN  & & OpenLANE \\
        \cline{2-4} 
                          & PSQ & PPA-Score & OpenLANE \\
    \end{tabular}
    \label{tab:metrics}
    \vspace{-5pt}
\end{table}

% \begin{table}[t]
%     \centering
%     \caption{SUBSTITUTED? Check the other table. Design goals and metrics for benchmarking RTL generation}
%     \begin{tabular}{lll}
%         \toprule
%         \textbf{Design Goals}        & \textbf{Evaluation Metrics}  & \textbf{Tool Used}                    \\
%         \midrule
%         Line-Level Syntactic and  &   Exact \& Fuzzy & Edit Distance~\cite{x}   \\
%         Contextual Accuracy    & Matching &  and Similarity   \\
%         \hline
%         Syntax Correctness    & pass@k & Icarus Verilog~\cite{iverilog}    \\
%         \hline
%         Functional Correctness   & pass@k & Icarus Verilog~\cite{iverilog}    \\
%         \hline
%         Synthesizability         & pass@k & Yosys~\cite{wolf2013yosys}        \\
%         \hline
%         Post-Synthesis Design Quality & PPA-score    & OpenROAD~\cite{ajayi2019openroad} \\
%         \bottomrule
%     \end{tabular}
%     \label{tab:metrics_old}
% \end{table}

\subsection{Measures and Metrics}\label{subsec:metrics}

%TODO: Diference between code completion and S2R might be related with the prompt (module header)
% Biggets drops is functinality
% Deepseek-R1 reasoning is very effective. But with higher response time... not suitable for SLC (Cristian and MIquel)
% Call for more benchmarks

To evaluate the five goals described above (LCA, STX, FNC, SYN and PSQ) we use different tools and measures tailored to each case. 
For LCA, prior work~\cite{allam2024rtl} has explored both exact and Fuzzy Matching (FM) techniques. Exact Matching (EM) determines whether the generated line precisely matches the reference HDL line in the dataset, while FM relies on edit distance and semantic similarity metrics. Our experimentation shows both metrics are strongly correlated, which is why we report only the more precise EM.
%This dual approach accounts for cases where functionally equivalent but syntactically different variations exist, ensuring a more flexible yet rigorous evaluation of LLM performance in line completion tasks.
%\gk{check RTL-Repo paper}
%\gk{the following three paragraphs are not structured as the other two for LCA and PSQ. I would state that for STX, FNC, and SYN goals, pass@k metric is used and here is the definition of pass@k.}
%EP: I tried to address this comment. The modified version follows.

For MC and S2R tasks, we set up an evaluation pipeline that sequentially tests STX, FNC, SYN, and PSQ.
STX is evaluated by compiling a design along with its testbench using Icarus Verilog~\cite{iverilog} and checking for errors. 
If no errors are issued at compile time, FNC is evaluated by running the simulation executable generated by the compiler and checking if the testbench passes. Functionally correct codes are tested for SYN and PSQ, by synthesizing the design with OpenLANE~\cite{shalan2020building}.
For the LCA, STX, FNC, and SYN goals, we report the same metric \passmetric{$1$}. This is a \passmetric{$k$} metric~\cite{kulal2019spoc,chen2021evaluating}, which measures the probability that at least one of $k$ generated solutions passes the corresponding test criteria. We set $k=1$ to focus on the realistic requirement of generating the correct result on the first try.

%For those designs that pass SYN, PSQ is computed by extracting post-synthesis PPA metrics from the OpenLANE~\cite{shalan2020building} reports. 

For designs that successfully pass SYN, we propose a novel evaluation metric called the PPA-score to assess post-synthesis design quality. This metric measures the performance of LLM-generated synthesizable code relative to a golden, human-crafted reference. The PPA-score is derived by extracting post-synthesis PPA metrics from the OpenLANE~\cite{shalan2020building} reports.
The designs are synthesized using the open-source SKY130A PDK~\cite{edwards2020google} and constrained with a $10ns$ delay, which was empirically chosen considering the complexity of the designs within the tested benchmarks and the selected technology node\footnote{Note that TuRTLe enables to change design constraints seamlessly through its configuration files.}. Notice the framework and the analysis results are technology and constraints independent since identical settings are applied to both human- and LLM-generated designs, and the main goal remains to compare LLM performance against human baselines.

\paragraph{PPA-Score} \label{subsec:PPA-Score}

Given a MC or S2R benchmark with $n$ problems, the \turtle{} framework generates $m$ candidate solutions for each. Then, each generation is processed sequentially through the evaluation pipeline: STX, FNC, SYN, PSQ. Each stage only processes results that passed the previous one, and failures in previous stages are reported as automatic fails in the next ones, creating a cascade score where $STX \ge FNC \ge SYN$. It could happen that $SYN < PSQ$ because we contemplate the possibility of having PSQ better than the human reference.

STX, FNC, and SYN are binary evaluations, pass or fail, which are aggregated using \passmetric{$k$}. However, PSQ requires not only to aggregate real values that have to be analyzed in comparison with the reference PPA of the golden solution, but also to take into account that models that produce more synthetizable code can be evaluated on a larger set of problems, thus increasing the challenge and confidence in the results obtained. To achieve this, we let $p_{i,j}$ represent the PPA metric (power, performance, or area) from the LLM for candidate $j$ of problem $i$. The PPA-score is defined as the average of scores for each generation (represented as $\hat{p}_{i,j}$) computed according to the following steps:

\begin{enumerate}
    \item Each $p_{i,j}$ is compared against the corresponding PPA value $g_i$ of the golden solution. For that, instead of aggregating $p_{i,j}$ we compute $p_{i,j}/g_i\in(0,+\infty)$.
    \item For generations that do not pass STX, FNC and SYN evaluations, $p_{i,j}$ cannot be computed. As a result, we set a failure value of $2\cdot g_i$ (\eg producing a design two times bigger than the human reference in the case of the area metric). This approach also clips to zero the score of designs that pass the previous evaluations but perform worse than this threshold.
    \item We flip the result so that the metric behaves as the rest of the goals (higher is better).
\end{enumerate}

The score $\hat{p}_{i,j}$ of generation $j$ for problem $i$ is defined as
\[
\hat{p}_{i,j} =
\begin{cases}
2 - \min(p_{i,j}/g_i,2) & \text{if $p_{i,j}$ exists} \\
0 & \text{otherwise}
\end{cases}
\]
which has the following interpretation:

\begin{itemize}
    \item $\hat{p}_{i,j} = 0$ : means that the generation has not passed the previous stages, or that it requires twice or more the area, the power, or the delay (performance), when compared to the human reference. 
    \item $\hat{p}_{i,j} = 1$ : are designs with an area, power, or performance equal to that of the human reference.
    \item $\hat{p}_{i,j} = 2$ : can only be obtained by chips which occupy no space, execute in no time, or consume no energy.
\end{itemize}

The final formula, considering all generations of an LLM for a given benchmark, is just the average of these scores:
\begin{equation}
x\text{-score} = \left[\frac{1}{n\cdot m} \sum\limits_{i=1}^n \sum\limits_{j=1}^m\hat{p}_{i,j}\right]\cdot 100 \; \; [\%],
\end{equation}

where $x\in\{\text{Power, Performance, Area}\}$. The PPA-score is reported as the average of the scores for these three metrics, as all of them are found to follow similar distributions. Note that due to the cascade effect previously explained, the PPA-Score compresses the information from all previous stages. It is thus the most representative value of model performance of those included in \turtle{}. 

Since the framework includes results across multiple benchmarks, these need to be aggregated. A weighted average is used, considering the significant differences in individual benchmarks sizes (see Table~\ref{tab:benchmark_comparison}).

%Finally, note that under the assumption that all generations are synthetizable and produce the same PPA as the golden solution (\ie the human baseline), the PPA-score would be 100\%. The theoretical upper bound is 200, but it is impossible to achieve in practice.

\begin{table}[t]
    \footnotesize
    \centering
    \renewcommand{\arraystretch}{1.2}
    \caption{Design goals covered by RTL benchmarks. Number of designs reports two sizes: Samples for LCA (left) and samples for the rest of goals (right).}
    \label{tab:benchmark_comparison}
    \begin{tabular}{l|c|c|c|c|c|c@{\hspace{0em}}c}
        \textbf{Benchmark} & \textbf{LCA} & \textbf{STX} & \textbf{FNC} & \textbf{SYN} & \textbf{PSQ} & \multicolumn{2}{c}{\textbf{Num. Designs}} \\
        %& & & & &  & (LCA) & (Rest) \\
        \hline
        RTL-Repo    & \checkmark & --         & --         & --         & --         & 1,174 &  -- \\
        VeriGen     & --         & \checkmark & \checkmark & \checkmark & --         &   --  & 17 \\
        VerilogEval & --         & \checkmark & \checkmark & \checkmark & --         &   --  & 156 \\
        RTLLM       & --         & --         & \checkmark & \checkmark & \checkmark &   --  & 50 \\
        \hline
        \turtle     & \checkmark & \checkmark & \checkmark & \checkmark & \checkmark & 1,174 & 223 \\
    \end{tabular}
\vspace{-10pt}
\end{table}

\subsection{Integrated Benchmarks}
\label{subsec:included_bench}

\turtle{} initially integrates a selection of four benchmarks, chosen by 
quality, variety and size. Table~\ref{tab:benchmark_comparison} provides a comparative overview of the design goals covered by each integrated benchmarks, together with the number of problem descriptions included. This number is reported separately for LCA (left) and the rest of goals (STX, FNC, SYN and PSQ, right), as they have different requirements.

\paragraph{RTL-Repo~\cite{allam2024rtl}} This benchmark is designed to evaluate LLMs' capabilities in the single-line completion task by assessing their ability to perform local edits within large-scale Verilog projects. It consists of 4,098 Verilog code samples sourced from public GitHub repositories. To construct prompts, multiple Verilog files from a given project are concatenated into a single input, which is then truncated at a predefined context length. The truncated prompt is fed to the LLM, requiring it to predict the next line of code immediately following the truncation point. Moreover, \turtle{} supports both the full dataset (4,098 samples) and the test split (1,174 samples) for benchmarking, but we adopt the test split as the default configuration. 
The test partition is large enough to produce stable results with minimal variance, reducing the risk of overfitting. More importantly, using only the test set mitigates data contamination risks, enhancing the credibility of the evaluation metric.
RTL-Repo primarily targets the LCA goal, which is the evaluation aspect implemented in \turtle{}.

%, and early experiments indicates test is large enough to produce stable results while being more efficient; using the full dataset provided an average gain of 4.33\% of exact match and 2.29\% of fuzzy match with standard deviations of 2.97\% and 1.15\% respectively. 

\paragraph{VeriGen~\cite{thakur2024verigen}} This benchmark evaluates LLMs' performance in the MC task using 17 Verilog problems categorized into basic, intermediate, and advanced difficulty levels.  Basic problems include simple components such as wires and logic gates, while advanced tasks involve more complex designs like finite state machines. For each problem, three levels of prompt specificity are provided: low, medium, and high with each offering different amounts of detail. The low-detail prompts include only the module header, which is generally insufficient for meaningful completion. In contrast, the high-detail prompts provide excessive information about the problem, making them impractical for real-world applications. To ensure a balanced and realistic evaluation, \turtle{} adopts the medium-detail prompts as the standard configuration. VeriGen was previously evaluated for STX, FNC, and SYN goals. In \turtle{}, we extend its scope by incorporating an additional design goal (PSQ) to assess the efficiency of LLM-generated RTL in terms of PPA.

\paragraph{VerilogEval~\cite{liu2023verilogeval,pinckney2025revisiting}}
This benchmark consists of 156 problems sourced from HDLBits\footnote{https://hdlbits.01xz.net/wiki/Problem\_sets}, designed to assess LLM's performance in two tasks: module completion (MC) and specification-to-RTL (S2R). In the MC task, models are provided with a problem statement along with a module header and are required to generate the missing body of the module. In the S2R task, models receive a prompt resembling a high-level design specification and must generate an entire module from scratch. %For evaluation, we use the default configurations for the hyperparameters as follows: a temperature of 0.2, top\_p of 0.95, a generation length of 2,048 tokens, and bf16 precision. 
%For models utilizing the reasoning token, we extend the generation length to 16,384 tokens to accommodate the additional reasoning steps.
VerilogEval was originally designed to evaluate STX, FNC and SYN design goals. \turtle{} extends its evaluation scope by incorporating PSQ. We manually corrected the reference implementation of six problems, to address synthesizability issues as explained in Appendix~\ref{sec:Appendix}.

\paragraph{RTLLM~\cite{lu2024rtllm}} This benchmark introduces 50 human-crafted designs of varying complexity, designed to evaluate LLMs on S2R. Originally, RTLLM was developed to assess LLM performance across three evaluation goals: syntax correctness (our SYN), functional correctness (our FNC), and design quality including power, performance, and area metrics (our PSQ).
Notice RTLLM definition of syntax correctness is differently from the STX goal used in this work. Specifically, RTLLM evaluates syntax correctness based on whether a design can be successfully synthesized into a netlist without syntax errors which is effectively combining aspects of both STX and SYN as defined in this work. 
%In this sense, \turtle{} provides a finer-grained assessment of LLM-generated Verilog.
RTLLM originally employs a ``success rate metric,'' which is conceptually similar to the \passmetric{$k$} metric~\cite{chen2021evaluating} used in this work, but it does not strictly adhere to the requirement that $N > k$.
%\turtle{} avoid potentially biased estimations by using $k=1$ and $N=5$ \manote{We already talk about this in III A. I would either reference it or remove it}.
%To integrate RTLLM into \turtle{}, we replace its commercial Synopsys VCS simulator with open-source alternatives, Icarus Verilog~\cite{iverilog} and OpenLANE~\cite{shalan2020building}, ensuring broader accessibility and reproducibility \manote{This tools apply to the other benchmarks, and it is already reflected in Table I}. 
%During this transition, all testbench results are manually reviewed, and one problematic design (\texttt{radix2\_div}, which failed to provide functional coverage) was excluded from evaluation \manote{Is this not referenced in the appendix Emanuele? I would reference it there and remove the explanation here, because we are not giving this level of detail for other benchmarks}. 
We corrected the reference implementation of four problems and excluded two to address functional and synthesizability issues as explained in Appendix~\ref{sec:Appendix}.

\section{Experiments}
\label{sec:eval}

%This section contains the analysis conducted and results obtained during experimentation. \S\ref{subsec:experimentalSetup} reviews the models benchmarked. From the results obtained, two sort of insights are gained. First, insights on the setting and reporting of results, as discussed in \S\ref{subsec:bench_insights}. This section defines and motivates the evaluation methodology implemented by \turtle. The second outcome is derived from the benchmarking of open models itself, using all five implemented goals. This provides a novel view on the current state-of-the-art, and findings on the nature of models and goals (see \S\ref{subsec:sota_benchmark}).

%\subsection{Generative Models}\label{subsec:models}
%A wide variety of open LLMs are assessed with the complete \turtle{} scope. This includes three types of models: General purpose LLMs (\eg LLaMA\cite{touvron2023llama}, DeepSeek\cite{zhu2024deepseek}) with good overall performance. Coding LLMs (\eg QwenCoder~\cite{hui2024qwen2} and  OpenCoder\cite{huang2024opencoder}) optimized specifically for code generation. And domain specific LLMs (\eg CodeV\cite{zhao2024codev}, HaVen\cite{yang2025haven}), tuned for RTL generation. Table ~\ref{tab:model_performance} shows results for models in these three categories.

This section presents the comprehensive evaluation of LLMs for RTL generation, analyzing their performance across multiple tasks and design goals using the \turtle{} framework. The following subsections detail the key findings and complete results. Models used are chosen to guarantee a diverse set of open LLMs, belonging to a wide range of categories. This includes general-purpose LLMs, such as LLaMA~\cite{touvron2023llama}, code-specific LLMs, such as OpenCoder~\cite{huang2024opencoder}, and RTL-specific LLMs, such as CodeV~\cite{zhao2024codev}.
In addition, we explore base models and instruct models, when both versions are available, such as Qwen~\cite{yang2024qwen2}. Finally, we consider the most recent family of models, trained to perform autoregressive reasoning chains before producing a final response commonly known as \textit{reasoning models}, such as DeepSeek R1~\cite{deepseekai2025deepseekr1}\footnote{Reasoning models are not tested for SLC, as their long reasoning chains are not adequate for a task that requires fast responses. Only Qwen3-A22B~\cite{qwen3technicalreport} SLC score is computed, as it has been fine-tuned to support both reasoning and non-reasoning modes.}. In total, our comparison includes forty LLMs, all executed under identical and reproducible conditions. Closed models are not included in this study, as these conditions could not be guaranteed.

\subsection{Benchmarking Insights}\label{subsec:bench_insights}

\begin{figure}[bt]
    \centering
    \includegraphics[width=\linewidth]{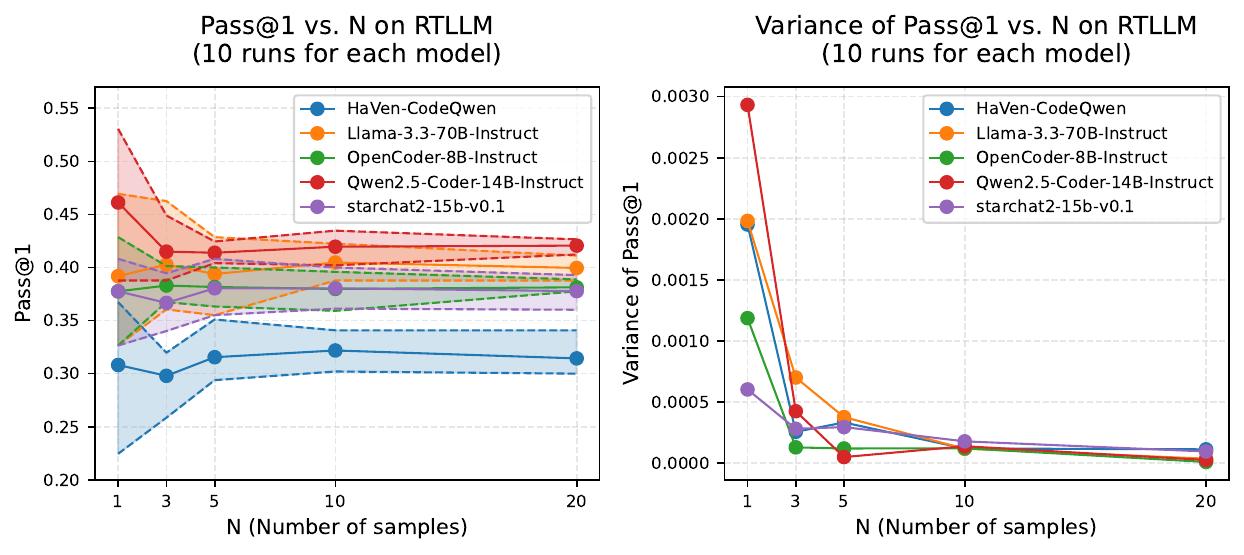}
    \caption{Preliminary study conducted on a subset of models, showing the \passmetric{$1$} variance among ten runs while increasing sample size $N$.}
    \label{fig:variance_N}
\vspace{-5pt}
\end{figure}

\paragraph{\passmetric{$1$} variance} 
We use the \passmetric{$1$} metric to measure the probability that a model generates a correct solution on the first attempt. However, the consistency of such metric is dependent on the number of samples ($N$) used per evaluation. To determine the optimal value of $N$, we performed ten independent runs on five selected models using varying sample sizes of $N={1, 3, 5, 10, 20}$. As shown in Figure \ref{fig:variance_N}, the variance decreases as $N$ increases, confirming that small sample sizes lead to unstable results. A sample size of $N=5$ is chosen as the optimal configuration, balancing computational efficiency with statistical stability.

\newcolumntype{C}[1]{>{\centering\arraybackslash}m{#1}} % Centered fixed-width column

\begin{figure*}[ht]
    \centering
    \includegraphics[width=\textwidth]{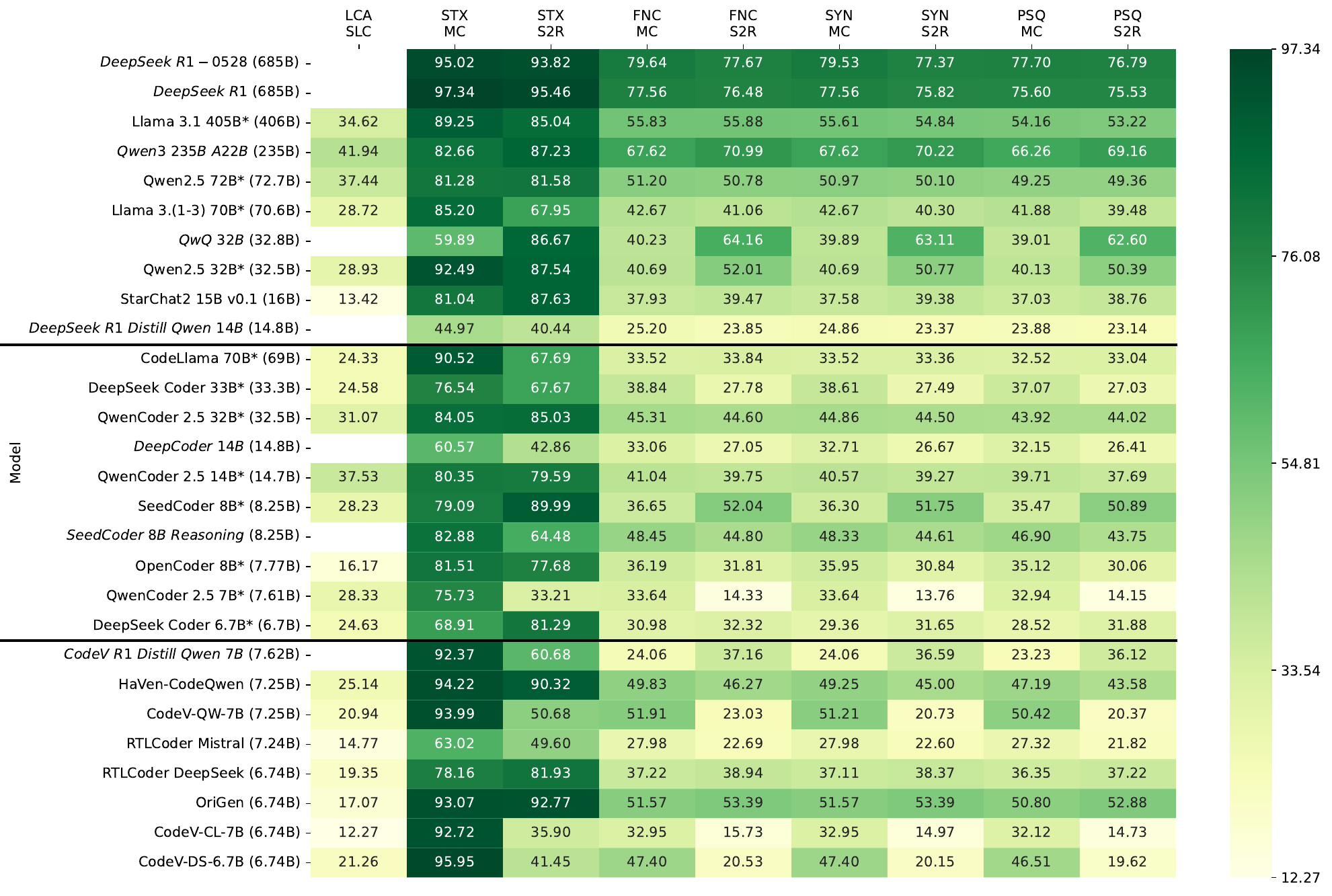}
    \caption{Pass@1 performance (higher is better) with n=5 for the five goals proposed. SLC: Single-Line Completion. MC: Module Completion. S2R: Specification-to-RTL. Table is split vertically in general purpose (top), coding (middle) and RTL-specific LLMs (bottom). Models marked with an asterisk (*) refer to the base version for SLC and MC tasks, and to the instruct version for S2R. The number in parentheses next to the model names indicates the model size, in billions of parameters. Reasoning models appear in italics. Reasoning models are not tested for SLC, except for Qwen3 235B A22B, which also supports non-reasoning mode.}
\label{fig:model_performance}
\end{figure*}

%\paragraph{Reducing variance of \smpassmetric{k}}
%Given \passmetric{1} (success at first attempt), we explore the best sampling size (N) for estimating the likelihood. Ten consecutive runs on 5 models across RTLLM are executed using N=\{1, 3, 5, 10, 20\}. Results, shown in Figure \ref{fig:variance_N} indicate variance decreases as the number of samples N grows. N=5 is fixed for all further experiments, as a trade-off between results variance and computational efficiency.

\paragraph{Temperature Ablation}

Prior to the main experiments, we examine the impact of stochasticity on LLM performance in RTL generation. Temperature controls the randomness of model-generated outputs, influencing both variance and precision. To assess its effect, we evaluated model performance across three temperature settings: 0.2, 0.5, and 0.8 for \passmetric{$1$}, focusing on STX and FNC goals, and affected by the choice of $N$ in the previous experiment in \ref{subsec:bench_insights}, we subsequently found that a temperature of 0.2 yielded the highest accuracy across all model categories (see Appendix~\ref{fig:temp_ablation}). This suggests that minimal randomness is optimal for RTL generation tasks. Consequently, 0.2 is set as the default temperature for all subsequent experiments to ensure consistency and reproducibility.

\paragraph{Single-Line Completion Context}
We analyze how the length of the context given in the prompt affects performance in single line completion tasks. We do this by evaluating six models, spanning general-purpose, coding and RTL-specific, using context sizes of 2,048, 4,096, and 8,192 tokens. Our results show that longer contexts consistently improve performance, confirming that LLMs benefit from additional context length when making predictions. Increasing the context length from 2,048 to 4,096 tokens results in a notable improvement of +3.35 in Exact Match scores with a standard deviation of 1.02. Extending the context further to 8,192 tokens provides an additional boost of +2.40 (standard deviation of 1.16). Based on these findings, 8,192 tokens is selected as the standard context length for the SLC task.

\paragraph{Base vs. Instruct-Tuned Models}  
\begin{figure}[bt]
    \centering
    \includegraphics[width=\linewidth]{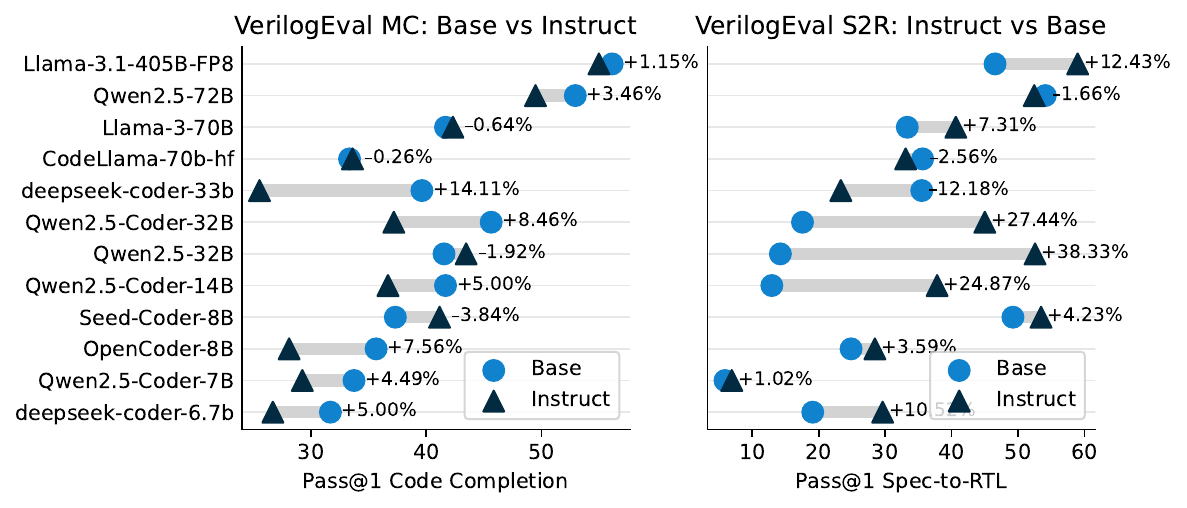}
    \caption{Comparison of Base and Instruct-Tuned model performance on VerilogEval for MC and S2R tasks across five model families.}
    %\caption{Performance of different models on the two variants of VerilogEval. x axis: Performance at MC. y axis: Performance at S2R. Diagonal: Equal performance.}
    \label{fig:base_ins_veval}
\vspace{-5pt}
\end{figure}
%VerilogEval v2 covers both module completion (MC) and specificication-to-RTL (S2R) tasks with the same underlying data. This provides an opportunity for assessing the impact of problem definition and presentation in model performance. Our experiments indicate base models are particularly appropriate for module completion (MC), while instruct-tuned models are typically best suited for the S2R task. Figure \ref{fig:base_ins_veval} illustrates this behavior, using the performance of the Base and Instruct variant on ten models, obtained from five different families (OpenCoder~\cite{huang2024opencoder}, Qwen 2.5~\cite{yang2024qwen2}, QwenCoder 2.5~\cite{hui2024qwen2}, DeepSeek Coder V1~\cite{zhu2024deepseek} and Llama 3.1~\cite{grattafiori2024llama}). %On code completion this holds for all models except for Llama 3.1 where the Instruct variant outperforms, in all cases, its base counter-part. On specification-to-RTL though we see how base models of OpenCoder 8B, QwenCoder 7B and DeepSeek Coder V1 33B outperform its instruct variant. 
%For the sake of computational efficiency and space constraints, we report Base variants on MC and SLC tasks, and Instruct variants on S2R tasks.

VerilogEval supports both MC and S2R tasks using the same underlying dataset, allowing for a direct comparison of how problem definition and presentation (\eg either a fill-in-the-blanks or an instruct-following task) impact model performance. This study is of special relevance for comparing base models against their instruct-tuned counterparts, and can help to optimize the selection and benchmarking of LLMs.

For those models where both versions are available, Figure~\ref{fig:base_ins_veval} compares the performance of base and instruct-tuned variants. That includes twelve models spanning six distinct model families: SeedCoder~\cite{seedcoder}, OpenCoder~\cite{huang2024opencoder}, Qwen 2.5~\cite{yang2024qwen2}, QwenCoder 2.5~\cite{hui2024qwen2}, DeepSeek Coder V1~\cite{zhu2024deepseek}, Llama 3~\cite{grattafiori2024llama} and CodeLlama~\cite{roziere2024codellama}. 
Results show a clear trend: base models tend to perform better in MC task, where they leverage learned syntax patterns to complete partial module definitions, whereas instruct-tuned models excel in S2R tasks, benefiting from their training on instruction-following tasks that improve their ability to interpret high-level specifications and follow them. The only notable exception to this comes from the \textit{deepseek-coder-33b} instruct model, which performs poorly on the S2R task. Further details on this experiment can be seen in Appendix~\ref{app:base-instr}. 

To maximize computational efficiency, for those models where both the base and the instruct version are available, we report base model performance on MC and SLC tasks, and the instruct-tuned models performance on S2R tasks.

%\paragraph{Context size in Line Matching}
%Previous work found counterproductive to use longer contexts in Line Matching~\cite{allam2024rtl}. 
%Our experiments indicate that providing longer contexts of the Verilog project to the LLM through the prompt increases performance in the Line Matching task. We evaluated six models (including general, coding, and domain-specific models of varying sizes and architectures) across three different context lengths (2,048, 4,096, and 8,192 tokens). In every case, performance improved with context length increased. Extending the context from 2,048 to 4,096 led to an average EM gain of 3.35 ($\sigma=1.02$), while increasing it to 8,192 provided +2.40 in EM ($\sigma=1.16$). The context length used in all further experimentation is 8,192 tokens.% Further experimentation could be conducted to see if this increasing trend continues with larger context sizes.

%The reason behind this might be in their implementation, as they are truncating the prompt from the right without effectively controlling its size, making it impossible to predict the next line for some samples.

\subsection{Model Results}\label{subsec:sota_benchmark}

%The evaluation provided by \turtle{} is comprehensive. An overview of the results are reported in Table~\ref{tab:model_performance}, while results for more models and additional views are made available through the public leaderboard released with this work. While the complete benchmark was conducted on 21 models, tables include only the best performing ones for all model categories (general purpose, coder and domain-specific).

This section discusses the performance results of LLM across the five design goals. See Figure~\ref{fig:model_performance} for the complete results, split by general purpose (top), coder LLMs (middle) and RTL-specific models (bottom).

\paragraph{Model Families} 
Models with include the autoregressive reasoning chain (DeepSeek R1-0528\cite{deepseekai2025deepseekr1}, DeepSeek R1 \cite{deepseekai2025deepseekr1}, Qwen3 235B A22B\cite{qwen3technicalreport}, QwQ 32B\cite{qwq32b}, DeepSeek R1 Distill Qwen 14B\cite{deepseekai2025deepseekr1}, DeepCoder 14B\cite{deepcoder2025}, SeedCoder 8B Reasoning\cite{seedcoder}, 'CodeV R1 Distill Qwen 7B\cite{CodeV-R1-Distill-Qwen-7B}) tend to be best-performing models across benchmarks. Their particularly competitive performance is also influenced by by their larger parameter count (\eg 685B for the two versions of DeepSeek-R1 and 235B for Qwen3-A22). However, the performance improvement on RTL design goals comes at the cost of increased computational overhead during inference, as these reasoning models produce substantially longer generations; All reasoning models were evaluated using a maximum context length of 16,384 tokens, making room for their reasoning chains, whereas the rest of models performed effectively with 2,048 tokens. Experiments to measure the impact of reasoning chain length on model performance are detailed in Appendix~\ref{app:reasoning_len}.
While large reasoning models achieve the highest scores, smaller RTL-specific models emerge as a particularly efficient solution for RTL generation. LLMs with either 6B or 7B parameters, such as Origen\cite{origen} and HaVen-CodeQwen\cite{haven}, demonstrate stable performance across multiple tasks, highlighting the importance of specialization for RTL tasks. This suggests two primary strategies for creating high-quality LLMs for RTL generation: either even larger general-purpose LLMs or small but highly specialized models fine-tuned using RTL data.

\begin{figure}[bt]
    \centering
    \includegraphics[width=\linewidth]{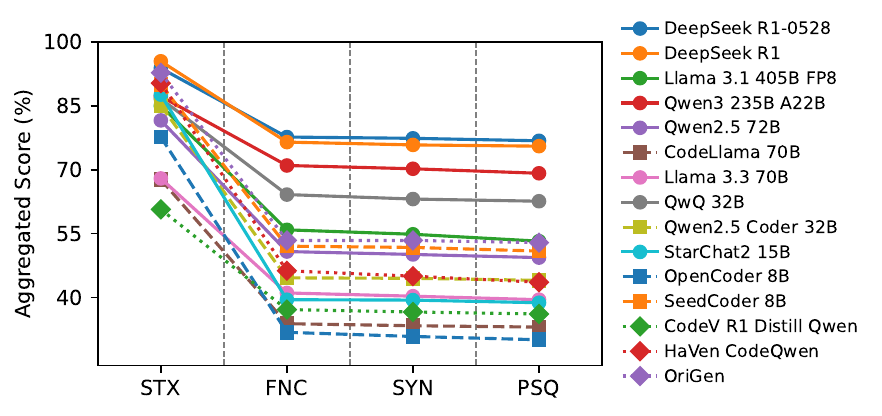}
    \caption{Aggregated model performance across design goals for specification-to-RTL task (VerilogEval \& RTLLM). General purpose models are shown with solid lines, coding models with dashed lines, and RTL-specific models with dotted lines.}
    \label{fig:goal_drops}
    \vspace{-5pt}
\end{figure}

%\begin{figure}[bt]
%    \centering
%    \includegraphics[width=0.9\linewidth]{images/drop_plot_top5.pdf}
%    \caption{Visualization of the sequential filtering for top 5 results in the area score of RTLLM.The dashed line represents the human baseline for each model.}
%    \label{fig:goal_drops_ppascore}
%\end{figure}

\paragraph{Model Performance by Design Goals}

% Figure~\ref{fig:goal_drops} shows average model performance across the design goals of STX, FNC, SYN, and PSQ for the specification-to-RTL task, evaluated across all benchmarks. Since these \turtle{} goals follow a cascading evaluation, the performance observed in this plot illustrates how RTL-generated code performance drops, as the evaluation criteria become progressively challenging.

According to the results of Figure~\ref{fig:model_performance}, the single-line completion task, measured through LCA, is the most challenging of all. This might be due to the existence of long-range flow, data dependencies, and multi-file contexts~\cite{allam2024rtl}. 
The rest of design goals can be analyzed visually with a cascade plot like the one in Figure~\ref{fig:goal_drops}, which shows average model performance across STX, FNC, SYN, and PSQ for the specification-to-RTL task, evaluated across all benchmarks. Since these \turtle{} goals follow a cascading evaluation, the performance observed in this plot illustrates how RTL-generated code performance drops, as the evaluation criteria become progressively challenging.
On average, models achieve a relatively high STX score of 76.36\%, indicating that generating syntactically correct Verilog is not a hard challenge in most cases. When, transitioning from syntax correctness to functional correctness results suffer a substantial drop, with the FNC score averaging a 42.37\% with a sharp decrease of -33.99\%. This decline suggests that while models are proficient at producing Verilog code that adheres to syntax rules, they struggle to generate functionally valid implementations. These results identify which should be the immediate next goal for LLMs in the context of RTL generation, and where should efforts be put during training efforts.
After FNC, it can be observed how all models perform very close to their human baseline, and no significant drops take place, neither at SYN nor at PSQ. This is likely influenced by the nature of the codes that do pass the FNC stage, which are likely to be the simplest ones, representing little challenge on the later SYN and PSQ stages (with some exceptions, see Appendix~\ref{app:low_psq}). Once LLMs manage to pass the FNC stage of the more challenging samples (roughly 25\% of the ones currently included in \turtle), the SYN and PSQ goals may exhibit significant drops.

\section{Conclusions}\label{sec:concl}

This work presents \turtle, the first unified evaluation framework for assessing LLMs in RTL generation, running across various tasks (single-line completion, module completion, and specification-to-RTL) and design goals (line-level contextual accuracy, syntax correctness, functional correctness, synthesizability, and post-synthesis quality). \turtle{} integrates multiple benchmarks into an automated and scalable pipeline that simplifies and standardizes experimentation. The framework is extendable by design both with new benchmarks and new models. To enhance the visibility and accessibility, an online \turtle{} leaderboard is released and maintained\footnote{\leaderboardurl}, as illustrated in Appendix~\ref{app:leaderboard}.

Results obtained from \turtle-based experimentation, the largest to date on RTL generation with forty models, reveal several critical insights. 
%\manote{start} LLMs benefit from longer context of the Verilog project for single line completion and lower temperatures improve precision \manote{end}. 
Reasoning-based models achieve state-of-the-art performance, despite their higher token generation requirements and inference latency. Model type also influences task performance, as base models outperform instruction-tuned models in completion tasks, while instruct-tuned models are more effective in specification-to-RTL tasks. Small and specialized models, tuned specifically with RTL data, perform competitively even with substantially fewer parameters.

Analysis of RTL tasks indicates LLMs approach saturation on current syntax benchmarks. However, functional correctness remains challenging, highlighting what should become the short-term research priority. Our contribution extends current assessment capabilities, by introducing a novel PPA score which quantifies post-synthesis RTL quality relative to human references. The proposed evaluation scheme acts as a cascade, linking advancements across core RTL generation requirements.

Overall, this work represents a significant step towards a comprehensive evaluation of LLMs for EDA workflows, providing insights into their capabilities and limitations in RTL generation. Our findings underscore the need for more sophisticated and realistic benchmarks that better reflect real-world design challenges, including multi-module architectures and complex module interconnections.
The \turtle{} framework source code is released on GitHub\footnote{\githuburl}, ensuring reproducibility. %A complete leaderboard, including more models and metrics, is hosted in HuggingFace\footnote{\leaderboardurl}.
Future work will expand this framework and its associated leaderboard with additional benchmarks, further supporting the evaluation and development of better LLMs for hardware design automation.

\section*{Acknowledgment}\label{sec:aknowledgment}
This work is supported by the AI4S fellowships awarded to Gokcen Kestor, Emanuele Parisi, Razine Moundir Ghorab, Cristian Gutierrez and Miquel Albertí Binimelis as part of the “Generación D” initiative, Red.es\footnote{\redesurl}, Ministerio para la Transformación Digital y de la Función Pública, for talent attraction (C005/24-ED CV1). Funded by the European Union NextGenerationEU funds, through PRTR. Additionally, this work has been partially funded by the Generalitat de Catalunya (contracts 2021-SGR-00763 and 2021-SGR-01187), and by the project PID2023-146511NB-I00 funded by the Spanish Ministry of Science, Innovation and Universities MCIU /AEI /10.13039/501100011033 and EU ERDF. We are grateful to the Operations department at BSC for their technical support.

% Include bibliography
\bibliographystyle{IEEEtran} % define style
\bibliography{references} % load file .bib

% Include appendix
\clearpage

\appendix{}
\label{sec:Appendix}
\subsection{FNC and SYN issues in RTLLM and VerilogEval}\label{app:bench_issues}

While deploying RTLLM and VerilogEval in the \turtle framework, we encountered some problems related to the human-crafted golden implementation of some of the samples that failed the FNC or SYN checks.
We encountered these problems in 6 samples from VerilogEval and 6 samples from RTLLM.
For each erroneous golden implementation, we adopted a coping strategy depending on the extent of changes required to manually fix the problem.
We decided to manually fix issues that could be fixed with limited human effort due to very clear problems in the sample code.
On the other hand, we decided to exclude samples whose fix would require more effort, or samples representing designs that are not synthesizable by construction.

\subsubsection{VerilogEval}

Problems \texttt{Prob95}, \texttt{Prob96}, \texttt{Prob137}, \texttt{Prob146}, \texttt{Prob152} and \texttt{Prob155} of VerilogEval fail the SYN check because Yosys infers a latch inside a \texttt{always\_comb} block.
This is caused by a non-exaustive \texttt{case} statement without a \texttt{default} clause.
Since all faulty examples represented finite-state machine designs where the \texttt{case} statement is used to compute the next state, we manually fixed these examples by adding the missing \texttt{default} branch, where the next FSM state is assigned a random state among the valid ones, depending on the problem specification.

\subsubsection{RTLLM}

The human-created solution of RTLLM problem \texttt{radix2\_div} does not meet the FNC goal, while \texttt{alu}, \texttt{multi\_booth\_8bit}, \texttt{clkgenerator}, \texttt{float\_multi} and \texttt{synchronizer} fail during synthesis or later checks in the OpenLANE classic flow.
Problem \texttt{radix2\_div} fails the FNC check when built with its own testbench using Icarus Verilog.
This problem does not occur with any other open-source or commercial simulator, so we excluded this sample from our evaluation.
Problem \texttt{alu} fails in the post-synthesis OpenLANE checks because the device under test has three undriven pins.
We solved this problem manually by commenting out the undriven pins in the module.
Problem \texttt{multi\_booth\_8bit} fails during synthesis because the \texttt{proc} Yosys command fails to translate all processes to netlist in the OpenLANE synthesis script. 
We have excluded this example from our evaluation.
Problem \texttt{clkgenerator} represents a clock generator module with one output port driven by a delayed statement.
While this example may be useful for evaluating the SYN and FNC goals, a clock generator is inherently not synthesizable, so we decided to exclude it from our evaluation.
Finally, \texttt{float\_multi} and \texttt{synchronizer} have bugs in some of their sensitivity lists, which we fixed by hand.

\subsection{Low PSQ of LLM-generated code: the case of \texttt{Prob30}} \label{app:low_psq}

During the PSQ analysis of some RTLLM and VerilogEval problems, we noticed that some of the LLM-generated patterns were functionally correct, but showed largely suboptimal PPA values with respect to the human-crafted golden solution.
It was particularly interesting what we observed for problem 30 from VerilogEval, which consists of computing the population counter of a 256-bit input vector. We observed across multiple generations of different models that LLM-generated answers reported much worse PPA metrics than the ones from the corresponding golden solution. Upon closer inspection, we saw that the bad generations looped over the 256-bit input array, including an unnecessary conditional to check if a bit was set before incrementing the population counter. 
The better implementation avoids the conditional statement altogether, and adds the bit directly to the counter, since if it is active, it can implicitly be the increment itself. 
While seemingly harmless, a conditional clause is synthesized into a multiplexer, and if the rest of the design contains only adder modules, adding 256 multiplexers has a massive impact on the PPA. 
This observation argues for a deeper investigation of LLM-generated code as future work to fully understand how to improve the PSQ of LLM-generated functional code to deliver design quality on par with what a well-trained RTL designer would do.

\subsection{Temperature ablation}\label{fig:temp_ablation}

\begin{figure}[bt]
    \centering
    \includegraphics[width=\linewidth]{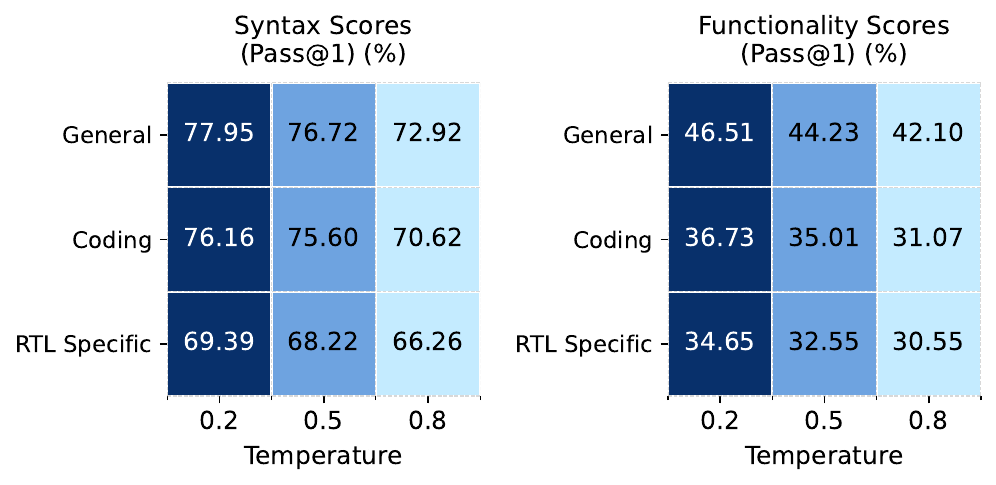}
    \caption{Pre-eliminary study done on 21 models. Average STX (left) and FNC (right) scores on VerilogEval and RTLLM across three model categories (General, Coding, and RTL-Specific) at different temperature settings (0.2, 0.5, 0.8). Darker shades indicate higher scores.}
\vspace{-5pt}
\label{fig:tmp}
\end{figure}

Figure~\ref{fig:tmp} shows model performance on different temperature settings, for the STX and FNC tasks. Results indicate optimal setting is temperature of 0.2.

\subsection{Quantitative evaluation of Base and Instruct model variants}\label{app:base-instr}

To quantitatively assess how the model variant affects performance, we start under the premise that base models, trained on next-token prediction, are better suited for pattern based completion benchmarks like VerilogEval's Module Completion (MC) benchmark. Similarly, we expect Instruct-tuned models to be better when provided with a set of instructions, as in Specification-to-RTL benchmarks.

In Table \ref{tab:mc_s2r_diff} we evaluate all models from our main results table that have both base and instruct variants on VerilogEval MC and S2R benchmarks. Initially,  8 out of 12 base models outperform its counter-variant on MC, while 9 out of 12 do the appropriate on S2R. To further quantify this impact we define $\Delta$ as the gap between benchmarks, this will tell us how balanced is the performance and towards which task is the model more inclined to perform better. Following on this, we perform the difference between both $\Delta_{\footnotesize{\textsc{B}}}$ and $\Delta_{\footnotesize{\textsc{I}}}$ denoted as:
\begin{equation*}
    \Delta\Delta = (\text{S2R}_{\textsc{INST.}} - \text{MC}_{\textsc{INST.}}) - (\text{S2R}_{\textsc{BASE}} - \text{MC}_{\textsc{BASE}})\;.
\end{equation*}

A positive $\Delta\Delta$ indicates that the instruct model variant generally perform better on S2R task relative to MC, and more so than the base variant. From the twelve models, eleven have a positive signed result. The only exception being CodeLlama 70B where although obtaining a close balance between model variants and benchmarks, it slightly inclines for the Base model performing better on S2R and the instruct variant being better at MC.

\begin{table}[h!]
\centering
\caption{Functionality (FNC) difference between Base and Instruct models on VerilogEval benchmarks: Module Completion (MC) and Spec-to-RTL (S2R).}
\label{tab:mc_s2r_diff}
\small
\begin{tabular}{@{}l | c@{\hskip 5pt}c | c@{\hskip 5pt}c | c@{\hskip 2pt}c@{\hskip 2pt} |@{\hskip 2pt} c@{\hskip -1pt}}
\toprule
\multirow{2}{*}{\textbf{Model}} 
  & \multicolumn{2}{c}{\textbf{MC}} 
  & \multicolumn{2}{c}{\textbf{S2R}} 
  & \multicolumn{2}{c}{\textbf{S2R$-$MC}} 
  & \multirow{2}{*}{$\Delta\Delta$} \\
  \cmidrule(lr){2-3} \cmidrule(lr){4-5} \cmidrule(lr){6-7}
  & Base & Ins. & Base & Ins. & $\Delta_{\footnotesize{\textsc{B}}}$ & $\Delta_{\footnotesize{\textsc{I}}}$ & \\
\midrule
  OpenCoder 8B & \underline{35.64} & 28.08 & 24.87 & \underline{28.46} & -10.77  & 0.38 & {\setlength{\fboxsep}{1.2pt}\colorbox{green!15}{\textbf{$+$}}} \\
Qwen2.5 C. 7B	& \underline{33.72} & 29.23 & 5.9 & \underline{6.92} & -27.82 & -22.31 & {\setlength{\fboxsep}{1.2pt}\colorbox{green!15}{\textbf{$+$}}} \\
CodeLlama 70B & 33.33 &	\underline{33.59} &	\underline{35.64} & 33.08 &	 2.31	& -0.51 & {\setlength{\fboxsep}{1.2pt}\colorbox{red!15}{\textbf{$-$}}} \\
DeepS. C. 6.7B & \underline{31.67} &	26.67 &	19.10 & \underline{29.62} &	-12.57	& 2.95  & {\setlength{\fboxsep}{1.2pt}\colorbox{green!15}{\textbf{$+$}}} \\
Qwen2.5 C. 32B & \underline{45.64} &	37.18 &	17.56 & \underline{45.00} &	-28.08	& 7.82  & {\setlength{\fboxsep}{1.2pt}\colorbox{green!15}{\textbf{$+$}}} \\
Llama 3 70B & 41.67 &	\underline{42.31} &	33.33 & \underline{40.64} &	-8.34	& -1.67 & {\setlength{\fboxsep}{1.2pt}\colorbox{green!15}{\textbf{$+$}}} \\
Qwen2.5 C. 14B & \underline{41.67} &	36.67 &	12.95 & \underline{37.82} &	-28.72	& 1.15  & {\setlength{\fboxsep}{1.2pt}\colorbox{green!15}{\textbf{$+$}}} \\
Qwen2.5 32B & 41.54 &	\underline{43.46} &	14.23 & \underline{52.56} &	-27.31	& 9.1   & {\setlength{\fboxsep}{1.2pt}\colorbox{green!15}{\textbf{$+$}}} \\
DeepS. C. 33B & \underline{39.62} &	25.51 &	\underline{35.51} & 23.33 &	-4.11	& -2.18 & {\setlength{\fboxsep}{1.2pt}\colorbox{green!15}{\textbf{$+$}}} \\
SeedCoder 8B & 37.31 &	\underline{41.15} &	49.23 & \underline{53.46} &	 11.92	& 12.31 & {\setlength{\fboxsep}{1.2pt}\colorbox{green!15}{\textbf{$+$}}} \\
Qwen2.5 72B & \underline{52.95} &	49.49 &	\underline{54.10} & 52.44 &	 1.15	& 2.95  & {\setlength{\fboxsep}{1.2pt}\colorbox{green!15}{\textbf{$+$}}} \\
Llama 3 405B & \underline{56.15} &	55.00 & 46.54 & \underline{58.97} &	-9.61	& 3.97  & {\setlength{\fboxsep}{1.2pt}\colorbox{green!15}
{\textbf{$+$}}} \\

\bottomrule
\end{tabular}
\end{table}

\subsection{Impact of reasoning CoT length}\label{app:reasoning_len}

Reasoning models have the particularity of generating an auto-regressive chain-of-thought where they prepend their reasoning enclosed into \texttt{<think>$\dots$</think>} tokens. Since the reasoning length cannot be directly controlled, we define an hyper-parameter, CoT length, to indicate the maximum number of tokens a model can allocate to produce the reasoning chain and final answer.

We select QwQ-32B, a top-performing reasoning model on the RTLLM benchmark.  In Table~\ref{tab:cot_length_qwq} we examine the impact of the CoT length by comparing the functionality score (pass@1) achieved on the RTLLM v2 benchmark, the percentage of truncated reasoning chains (cases where the model exhausts the token window likely during the reasoning phase, failing to produce a final answer), and the corresponding inference times.

All experiments were conducted in a single node with 4x Nvidia Hopper GPU with 64 HBM2 memory.

\begin{table}[h!]
\centering
\caption{Impact of CoT length on QwQ-32B (RTLLM benchmark).}
\small
\label{tab:cot_length_qwq}
\begin{tabular}{@{}lccc@{}}
\toprule
\textbf{CoT length} & \makecell{\textbf{FNC}\\\textbf{(pass@1)}} & \makecell{\textbf{Truncated}\\\textbf{CoTs}} & \makecell{\textbf{Inference Time}\\\textbf{(mm:ss)}} \\
\midrule
8K                   & 46.53\%               & 42.45\%           & 14:01 \\
16K                  & 56.32\%      & 5.31\%            & 24:32 \\
32K (recommended)                  & 55.91\%               & 0.41\%            & 27:44 \\
\bottomrule
\end{tabular}
\end{table}

From the results we see how the shortest window of 8K tokens, leads to poor performance due to high truncation rates, almost half of its generations were cut at the middle of either the reasoning phase or the final code answer. Increasing the CoT length to 16K suposes a big increase in functionality (pass@1) with 5.31\% of the samples being truncated. Surprisingly, the 16K window outperforms slightly the recommended by the model provider (32K), we attribute this phenomenon the the model-agnostic post-processing, where it will parse the latest produced code regardless. This suggests that QwQ often produces valid answers before completing the full reasoning sequence. Finally, we see how the inference time from 8K to 16K almost doubles, while from 16K to 32K grows more modestly. This aligns with the number of truncated generations, where 5.31\% exceed the 16K token window.

\subsection{Leaderboard HuggingFace space}\label{app:leaderboard}

\begin{figure}[h!]
  \centering
  \caption{Screenshot of the \turtle{} leaderboard.}
  \includegraphics[width=0.93\textwidth]{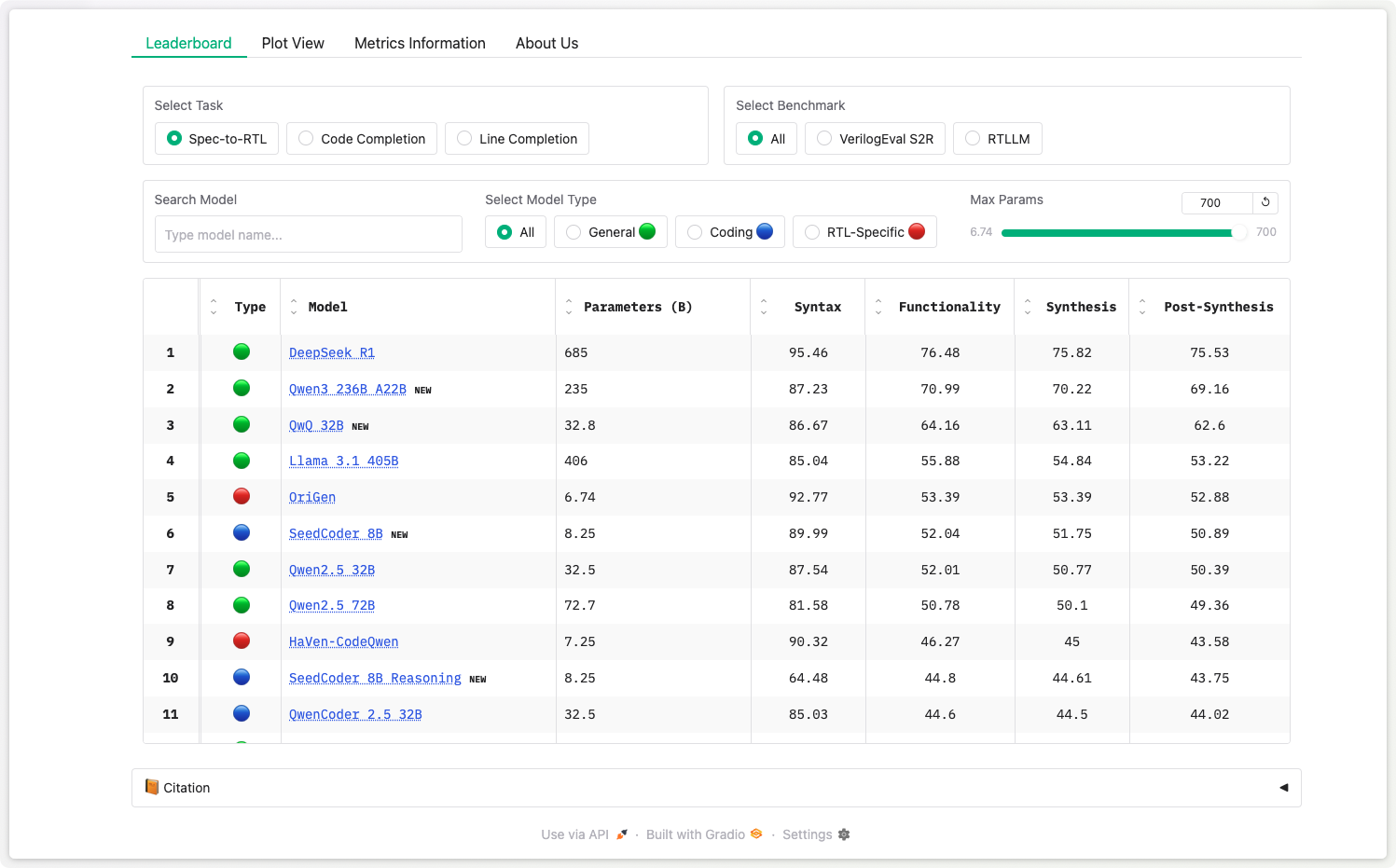}
  \label{fig:leaderboard_screenshot}
\end{figure}

\begin{figure}[h!]
  \centering
  \caption{Screenshot of the \turtle{} plot view.}
  \includegraphics[width=0.93\textwidth]{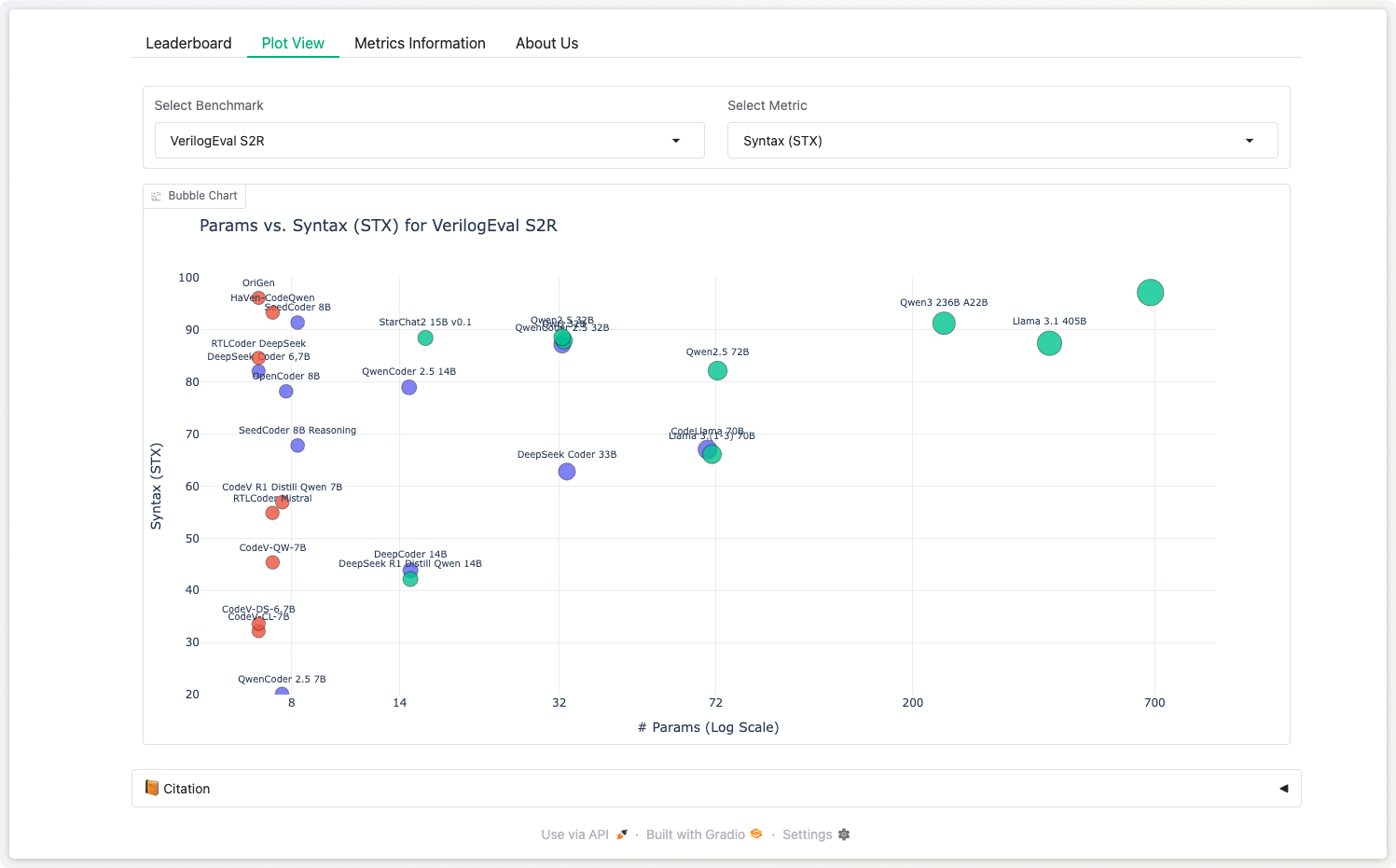}
  \label{fig:leaderboard_screenshot_2}
\end{figure}

Figure~\ref{fig:leaderboard_screenshot} shows the leaderboard released with this work, including over thirty LLMs, and all the visualization options available (task, benchmark, model type and number of parameters). The same data is shown as a plot using the number of parameters in the x axis in Figure~\ref{fig:leaderboard_screenshot_2}, which facilitates the visualization and interpretation of trends across model families and scales.

\onecolumn

\end{document}